\documentclass[a4paper,11pt]{article}
\usepackage[left=2cm,right=2cm,top=2cm,bottom=2cm]{geometry}
\usepackage{psfrag,amsmath,amsfonts,amssymb,latexsym,amsthm,verbatim,color,lscape}
\usepackage[figuresright]{rotating}
\usepackage{moreverb}

\usepackage{mathtools}

\usepackage{multicol}
\usepackage{amsmath,lscape,amsfonts,amssymb,
amsthm,verbatim,color,lscape,psfrag}

\usepackage{url}
\usepackage{hyperref}
\hypersetup{colorlinks,
citecolor=black,
filecolor=black,
linkcolor=black,
urlcolor=black,
pdftex}

\usepackage{amsmath}
\usepackage{algorithm}
\usepackage[noend]{algpseudocode}

\usepackage[compress]{natbib}
\usepackage{booktabs}

\usepackage{colortbl}
\definecolor{gray}{rgb}{0.8,0.8,0.8}

\usepackage{times}
\usepackage{blindtext}
\usepackage[font={footnotesize,it}]{caption}

\usepackage{sectsty}  
\subsectionfont{\normalsize\bf}
\sectionfont{\large\bf}

\def \red#1{\textcolor{red}{#1}}

\def \red#1{\textcolor{red}{#1}}

\begin{document}

\def\p{\partial}
\def\oo{\infty}
\def\rt#1{\sqrt{#1}\,}

\def\Cbar{{\overline C}}
\def\C{\mathbf{C}}
\def\E{{\rm E}\,}
\def\I{\mathbf{I}}
\def\pp{\mathbf{p}}
\def\R{\mathbf{R}}
\def\y{\mathbf{y}}
\def\Y{\mathbf{Y}}
\def\z{\mathbf{z}}
\def\x{\mathbf{x}}
\def\o{\omega}
\def\s{\sigma}

\def\V{\mathbf{V}}
\def\I{\mathbf{I}}
\def\bfv{\mathbf{v}}
\def\X{\mathbf{X}}
\def\D{\mathbf{D}}

\def\a{{\alpha}}
\def\g{\gamma}
\def\b{\beta}

\def\de{\delta}
\def\debf{\boldsymbol{\delta}}
\def\gbf{\boldsymbol{\gamma}}
\def\e{\epsilon}
\def\th{\theta}
\def\r{\rho}
\def\thbf{\boldsymbol{\theta}}
\def\taubf{\boldsymbol{\tau}}
\def\pibf{\boldsymbol{\pi}}
\def\Xibf{\boldsymbol{\Xi}}
\def\Sbf{\boldsymbol{\Sigma}}
\def\sbf{\boldsymbol{\sigma}}

\def \red#1{\textcolor{red}{#1}}
\def \blue#1{\textcolor{blue}{#1}}
\def \magenta#1{\textcolor{magenta}{#1}}
\def \green#1{\textcolor{green}{#1}}
\def\bbf{\boldsymbol{\beta}}
\def\bmu{\boldsymbol{\mu}}

\title{A multinomial truncated D-vine copula mixed model for  the joint meta-analysis of multiple diagnostic tests}

\date{}
\author{
Aristidis K. Nikoloulopoulos\footnote{{\small\texttt{A.Nikoloulopoulos@uea.ac.uk}}, School of Computing Sciences, University of East Anglia, Norwich NR4 7TJ, UK} }
\maketitle

\begin{abstract}
\baselineskip=24pt
\noindent
There is an  extensive  literature on  methods for meta-analysis of diagnostic studies, but it 
 mainly focuses on a single test. 
However,  the better  understanding of a particular disease has led to the development of multiple tests. A multinomial generalized linear mixed model (GLMM) is recently proposed for the joint meta-analysis of studies comparing multiple tests. 
We propose a novel model for the joint meta-analysis of multiple tests, which assumes independent multinomial distributions for the counts of each combination of test results in diseased and non-diseased patients, conditional on the latent vector of probabilities of each combination of test results in diseased and non-diseased patients.  For the random effects distribution of the latent proportions, we employ a truncated drawable vine copula that  can cover flexible dependence structures. The proposed model includes the multinomial GLMM as a special case,  but can also operate on the original scale of the latent proportions. Our methodology is demonstrated with a simulation study and using a meta-analysis of screening for Down syndrome with two tests: shortened humerus and shortened femur. The comparison of our method with the multinomial GLMM yields findings in the real data meta-analysis that change the current conclusions.  

\end{abstract}

\baselineskip=24pt

\section{Introduction}
 Diagnostic test accuracy studies aim to identify a new diagnostic test that is as or more accurate than the current gold standard, yet less expensive or invasive. 
The  development of an accurate diagnostic test  leads to early identification of a particular disease and  substantially  reduces the healthcare costs.   
The  large  number of  available diagnostic test accuracy  studies has led to the use of meta-analysis as an integrated analysis that will have more statistical power to detect an accurate diagnostic test than an analysis based on a single study. As the accuracy of a diagnostic test
is commonly measured by a pair of indices such as sensitivity and specificity, synthesis of diagnostic test accuracy studies actually requires  multivariate meta-analysis methods, see e.g.,  \cite{JacksonRileyWhite2011}. 

There is an  extensive  literature on  methods for meta-analysis of diagnostic studies, but it 
 mainly focuses on a single test (e.g., \citealt{Reitsma-etal-2005,Chu&Cole2006,
RutterGatsonis2001,Nikoloulopoulos2015b}). 
However,  the better  understanding of a particular disease, along with the technological advance in many health sectors has led to the development of multiple tests.  
In the research area of detecting fetuses with Down syndrome, that motivated this work,   many  screening accuracy of second-trimester ultrasound markers have been developed. 
Down syndrome is the most common clinical significant chromosomal abnormality among fetuses (e.g., \citealt{Kliegman-etal-2007}). 
There has been a substantial interest in the prenatal detection of affected fetuses so that parents can be prepared for the birth of an affected child or even consider pregnancy  termination (e.g., \citealt{Smith-Bindman-etal-2001}). 
Mothers and fetuses identified by a positive screening test result are typically offered a definitive diagnosis via amniocentesis, an invasive diagnostic test (e.g., \citealt{trikalinos-etal-2014-rsm}).  

Hence, as the meta-analysis of  more than one diagnostic tests can clearly  impact clinical decision making and patient health,
\cite{trikalinos-etal-2014-rsm}  
proposed a multinomial generalized linear mixed model (GLMM) for the joint meta-analysis of two such markers, namely shortened humerus (arm bone) and shortened femur (thigh bone) of the fetus that screen for Down syndrome. Their model assumes independent multinomial distributions for the counts of each combination of test results in diseased patients, and, the counts of each combination of test results in non-diseased patients, conditional on the 6-variate normally distributed transformed latent  true positive rate (TPR) and false positive rate (FPR) for each test, and latent joint TPR and  FPR, which capture information on the agreement between the two tests  in each study.

Nevertheless,  the 6-variate normal  distribution of the transformed latent proportions in the multinomial GLMM  has restricted properties, i.e., a linear correlation structure and  normal margins that might lead to biased meta-analytic estimates of diagnostic test accuracy. 
In order to create a flexible distribution to model  the random effects we exploit the use of copula functions \citep{joe97,joe2014}.  We propose a multinomial  
copula mixed model (CMM) as an extension of the multinomial GLMM  by rather using a  copula representation of the random effects distribution with normal and beta margins.  
We  assume independent multinomial distributions for the counts of each combination of test results in diseased patients, and, the counts of each combination of test results in non-diseased patients, conditional on the latent probabilities of each combination of test results in diseased and non-diseased patients in each study.  We consider the case where the same individuals receive both tests and the results are cross-classified.

For the random effects distribution, the choice of the copula couldn't be other than the class of regular vine copulas \citep{Bedford&Cooke02} as other copulas such as Archimedean, nested Archimedean and elliptical copulas have limited dependence (see, e.g., \citealt{Nikoloulopoulos2013a}).  
Regular vine copulas are suitable for high-dimensional data (e.g., \citealt{Schepsmeier&Czado2016-jrssc}), 
hence given the low dimension $d=6$, 
we use their boundary case, namely   a drawable vine (D-vine) copula. D-vine copulas  have become important in  many applications areas such as finance \citep{aasetal09,nikoloulopoulos&joe&li11}
and biological sciences \citep{Killiches&Czado-2018,nikoloulopoulos-2018-smmr}, to just name a few,  in order to deal with dependence in the joint tails. Another boundary case of regular vine copulas is the canonical vine copula, but this parametric family of copulas is suitable if there exists a  variable that drives the dependence among the variables  \citep{nikoloulopoulos&joe12,Erhardt&Czado2018-JRSSC},  which apparently is not the case in this application area.

A $6$-dimensional D-vine copula
can cover flexible dependence structures, different from assuming simple linear correlation structures, tail independence and normality \citep{joeetal10}, through the specification
of five bivariate marginal copulas at level 1 and ten
bivariate conditional copulas at higher levels; at level $\ell$  for
$\ell=2,\ldots,5$, there are $6-\ell$ bivariate conditional copulas
that condition on $\ell-1$ variables. 
\cite{joeetal10} have shown that in order for a vine copula to have (tail) dependence for all bivariate margins, it is only necessary for the bivariate copulas in level 1 to have (tail) dependence and it is not necessary for the conditional bivariate copulas in levels $2,\ldots,5$ to have tail dependence. That provides 
some  theoretical justification for the idea to model the  dependence in the first level and then just use the independence copulas at higher levels without sacrificing the tail dependence of the vine copula distribution. 
This truncation or simplification, as per the terminology in \cite{Brechmann-Czado-Aas-2012}, offers   a substantial  reduction of the dependence parameters. For example, this  enabled \cite{Coblenz-etal-2020-jrssc}   to provide a parsimonious 
parametric model for the dependence structure of fuel drops generated by a fuel injector in a jet engine. In our case there are 10 dependence  parameters less, which  is extremely useful for estimation purposes given the typically small number of primary studies involved in the meta-analysis. 
The multinomial  truncated D-vine CMM    (a) includes the multinomial GLMM proposed by \cite{trikalinos-etal-2014-rsm} as a special case (b) can have arbitrary univariate distributions for the random effects, (c) can  provide tail dependencies and asymmetries and (d) has computational feasibility.

The remainder of the paper proceeds as follows. 
Section \ref{model} introduces the multinomial truncated D-vine CMM for meta-analysis of two diagnostic tests, and provides computational details for maximum likelihood (ML) estimation.  Section \ref{simulations-section} studies the small-sample efficiency of the proposed ML estimation technique,  investigates the effect of misspecifying the univariate margins of the random effects distribution on parameter estimators and standard errors, and compares the proposed methodology with the multinomial GLMM.
Section \ref{application} demonstrates our methodology by insightfully re-analysing the data  from a systematic review that examined the screening accuracy of two second-trimester ultrasonographic tests that screen for Down syndrome.
We conclude with some discussion in Section \ref{discussion}, followed by a brief section with  software details.

\section{\label{model}The multinomial  truncated D-vine copula mixed model}

In this section, we  introduce the  multinomial truncated D-vine CMM for the joint meta-analysis of two diagnostic tests.  In Subsections \ref{norm-model} and \ref{beta-model}, a truncated D-vine copula representation of the random effects distribution with normal and beta margins, respectively, is presented. 
We complete this section with details on ML estimation.

\subsection{\label{norm-model}The multinomial   truncated D-vine copula mixed model with normal margins}
 We first introduce the notation used in this paper. The data are  $y_{ijkt},\, i = 1, . . . ,N,\, j=0,1,\,k=0,1,\,t=0,1$, where $i$ is an
index for the individual studies, $j$ is an index for the test 1 outcome (0:negative; 1:positive), $k$ is an index for the test 2 outcome (0:negative; 1:positive)  and $t$  is an index for the disease outcome (0: non-diseased; 1: diseased). 
The ``classic" $2\times 2$ table  is extended to a $4\times 2$ table (Table \ref{4times2}). 
Each cell in  Table \ref{4times2} provides  the cell frequency 
corresponding to a combination of index tests and disease outcome in study $i$.

\begin{table}[!h]
\caption{\label{4times2} Data  from an individual study in a  $4\times 2$  table. }
\centering
\setlength{\tabcolsep}{46pt}
\begin{tabular}{ccccc}
\toprule
&&\multicolumn{2}{c}{Disease (by gold standard)}\\ 
{Test 1}& Test 2   & $+$ & $-$\\\hline
$+$ &$-$&$y_{i101}$& $y_{i100}$ \\
$-$ &$+$ & $y_{i011}$&  $y_{i010}$ \\
$+$ &$+$&$y_{i111}$& $y_{i110}$\\
$-$ &$-$&$y_{i001}$ &$y_{i000}$  \\
\hline
\multicolumn{2}{c}{Total} &  $y_{i++1}$&$y_{i++0}$\\
\bottomrule
\end{tabular}

\end{table}

We assume that  the counts of each combination of test results in diseased patients are 
multinomially distributed given 
$\X_1=\x_1$, where $\X_1=(X_{101},X_{011},X_{111})$ is the   latent vector of transformed  probabilities of each combination of test results in diseased patients. That is
\begin{multline}\label{within-diseased}
(Y_{i101},Y_{i011},Y_{i111},Y_{i001}
)|(X_{101}=x_{101},X_{011}=x_{011},X_{111}=x_{111})\sim\\ \mathcal{M}_4\Bigl(y_{i++1},
l^{-1}(x_{101},x_{011},x_{111}),
l^{-1}(x_{011},x_{101},x_{111}),
l^{-1}(x_{111},x_{101},x_{011})
\Bigr),\footnotemark
\end{multline}
\footnotetext{$\mathcal{M}_T(n,$ $p_1, \dots,p_{T-1})$  is shorthand notation for the  multinomial distribution; $T$ is the number of cells, $n$ is the number of observations, and  $(p_1,\dots,p_T)$ with $p_T = 1-p_1-\ldots-p_{T-1}$ is the $T$-dimensional vector of success probabilities.}
where  
$l^{-1}$  is the inverse multinomial logit link, e.g., $l^{-1}(x_{101},x_{011},x_{111})=\frac{e^{x_{101}}}{1+e^{x_{101}}+e^{x_{011}}+e^{x_{111}}}$.

In a similar manner we assume that 
the counts of each combination of test results in non-diseased patients  are  multinomially distributed given 
$\X_0=\x_0$, where $\X_0=(X_{100},X_{010},X_{110})$ is the   latent vector of transformed  probabilities of each combination of test results in non-diseased patients. That is
\begin{multline}\label{within-nondiseased}
(Y_{i100},Y_{i010},
Y_{i110},Y_{i000})|(X_{100}=x_{100},X_{010}=x_{010},X_{110}=x_{110})\sim\\ \mathcal{M}_4\Bigl(y_{i++0},
l^{-1}(x_{100},x_{010},x_{110}),
l^{-1}(x_{010},x_{100},x_{110}),
l^{-1}(x_{110},x_{100},x_{010})
\Bigr).
\end{multline}

After defining the within-studies model in (\ref{within-diseased}) and (\ref{within-nondiseased}), we next define the between-studies model. 
The stochastic representation of the between studies model takes the form

\begin{small}
\begin{align}\label{copula-between-norm}
&\biggl(\Phi\Bigl(X_{101};l(\pi_{101},\pi_{011},\pi_{111}),\s_{101}^2\Bigr),\Phi\Bigl(X_{011};l(\pi_{011},\pi_{101},\pi_{111}),\s_{011}^2\Bigr),\Phi\Bigl(X_{111};l(\pi_{111},\pi_{101},\pi_{011}),\s_{111}^2\Bigr),\\&\Phi\Bigl(X_{100};l(\pi_{100},\pi_{010},\pi_{110}),\s_{100}^2\Bigr),\Phi\Bigl(X_{010},l(\pi_{010},\pi_{100},\pi_{110}),\s_{010}^2\Bigr),\Phi\Bigl(X_{110};l(\pi_{110},\pi_{100},\pi_{010}),\s_{110}^2\Bigr)\biggr)\sim C_6(\cdot;\thbf),\nonumber
\end{align}
\end{small}
\hspace{-1.5ex} where $C_6(\cdot;\thbf)$ is a 6-dimensional truncated   D-vine  copula with dependence parameter vector 
$\thbf=(\th_{101,011},$ 
$\th_{011,111}, 
\th_{111,100},\th_{100,010}, \th_{010,110})$, $\Phi$  
is the cumulative distribution function (cdf) of the  N($\mu,\s^2$) distribution, and $l$  is the multinomial logit link, e.g., $l(\pi_{101},\pi_{011},\pi_{111})=\log\Bigl(\frac{\pi_{101}}{1-\pi_{101}-\pi_{011}-\pi_{111}}\Bigr)$. The copula parameter vector $\thbf$ has parameters of the random effects model and they are separated from the univariate parameter vectors $\pibf_t=(\pi_{10t},\pi_{01t},\pi_{11t})$  and $\sbf_t^2=(\s_{10t}^2,\s_{01t}^2,\s_{11t}^2)$. 
The $\pibf_t$'s  have the  actual parameters of interest, since $\pi_{111}$ and $\pi_{110}$ is the meta-analytic parameter of joint TPR  and joint FPR, respectively,  and the meta-analytic parameters for the TPR and  FPR in each test  are functions of these parameters, viz. 
\begin{equation}\label{TPR}
\pi_{1\cdot1}=\pi_{101}+\pi_{111}
\qquad \pi_{\cdot11}=\pi_{011}+\pi_{111}
\end{equation}
and 
\begin{equation}\label{FPR}
\pi_{1\cdot0}=\pi_{100}+\pi_{110}
\qquad \pi_{\cdot10}=\pi_{010}+\pi_{110},
\end{equation}
respectively. 
The univariate parameter vectors $\sbf_t^2$'s   denote the variabilities of the random effects.

The models in (\ref{within-diseased}--\ref{copula-between-norm}) together specify a multinomial truncated D-vine CMM with joint likelihood
\begin{align*}
&L(\pibf_1,\pibf_0,\sbf_1^2,\sbf_0^2,\thbf|\y_1,\y_0)
=\\
&\prod_{i=1}^N\int_{(-\infty,\infty)^6}
g\Bigl(y_{i101},y_{i011},
y_{i111};y_{i++1},l^{-1}(x_{101},x_{011},x_{111}),
l^{-1}(x_{011},x_{101},x_{111}),
l^{-1}(x_{111},x_{101},x_{011})\Bigr)\times\\
&g\Bigl(y_{i100},y_{i010},
y_{i110};y_{i++0},l^{-1}(x_{100},x_{010},x_{110}),
l^{-1}(x_{010},x_{100},x_{110}),
l^{-1}(x_{110},x_{100},x_{010})\Bigr)\times\\
&f(\x_1,\x_0;\pibf_1,\pibf_0,\sbf_1^2,\sbf_0^2,\thbf)d\x_1d\x_0, 
\end{align*}
where  $g(;n,p_1,\ldots,p_{T-1})$ is the $\mathcal{M}_T(n,p_1,\dots,p_{T-1})$ probability mass function (pmf) and  $f$ 
is the 6-variate truncated D-vine density, viz.
\begin{multline}\label{vine-density}
f(\x_1,\x_0;\pibf_1,\pibf_0,\sbf_1^2,\sbf_0^2,\thbf)=\\
\phi(x_{101})\phi(x_{011})\phi(x_{111})\phi(x_{100})\phi(x_{010})\phi(x_{110})
c_{6}\Bigl(\Phi(x_{101}),\Phi(x_{011}),\Phi(x_{111}),\Phi(x_{100}),\Phi(x_{010}),\Phi(x_{110});\thbf
),
\end{multline}
with 
\begin{align*}
&c_{6}\Bigl(\Phi(x_{101}),\Phi(x_{011}),\Phi(x_{111}),\Phi(x_{100}),\Phi(x_{010}),\Phi(x_{110});\thbf\Bigr)=
c_2\Bigl(\Phi(x_{101}),\Phi(x_{011});\th_{101,011}\Bigr)c_2\Bigl(\Phi(x_{011}),\\&\Phi(x_{111});\th_{011,111}\Bigr)
c_2\Bigl(\Phi(x_{111}),\Phi(x_{100});\th_{111,100}\Bigr)c_2\Bigl(\Phi(x_{100}),\Phi(x_{010});\th_{100,010}\Bigr)
c_2\Bigl(\Phi(x_{010}),\Phi(x_{110});\th_{010,110}\Bigr),
\end{align*}
where, e.g., $\phi(x_{101})$ and $\Phi(x_{101})$ is shorthand notation for  the density $\phi\bigl(x_{101};l(\pi_{101},\pi_{011},\pi_{111}),\s_{101}^2\bigr)$ and cdf  $\Phi\bigl(x_{101};l(\pi_{101},\pi_{011},\pi_{111}),\s_{101}^2\bigr)$ of the $N(\mu,\s^2)$ distribution, and $c_2$'s are bivariate copula densities. 
Note that for a 6-dimensional D-vine copula density there are $\frac{6!}{2}$ distinct decompositions \citep{aasetal09}. 
To be concrete in the exposition of the theory, we use the  decomposition in (\ref{vine-density}); the theory though also apply to the other  decompositions.

 Below we transform the original integral into an integral over a unit hypercube using the inversion method. Hence the joint likelihood becomes
\begin{align*}
&\prod_{i=1}^N\int_{[0,1]^6}
g\Bigl(y_{i101},y_{i011},
y_{i111};y_{i++1},l^{-1}(x_{101},x_{011},x_{111}),
l^{-1}(x_{011},x_{101},x_{111}),
l^{-1}(x_{111},x_{101},x_{011})\Bigr)\times\\
&g\Bigl(y_{i100},y_{i010},
y_{i110};y_{i++0},l^{-1}(x_{100},x_{010},x_{110}),
l^{-1}(x_{010},x_{100},x_{110}),
l^{-1}(x_{110},x_{100},x_{010})\Bigr)\times\\
&c_6(u_{101},u_{011},u_{111},
u_{100},u_{010},u_{110};\thbf)du_{101}du_{011}du_{111}du_{100}du_{010}du_{110}, 
\end{align*}
where, e.g., $x_{101}=\Phi^{-1}\Bigl(u_{101};l(\pi_{101},\pi_{011},\pi_{111}),\s_{101}^2\Bigr)$.

\subsubsection{Relationship with the multinomial GLMM}
In this section, we show what happens when all the bivariate copulas are bivariate normal (BVN) and the univariate distribution of the random effects is the $N(\mu,\s^2)$ distribution. We can easily deduce that the within- study model in  (\ref{within-diseased}) and (\ref{within-nondiseased}) is the same as in the multinomial  GLMM \citep{trikalinos-etal-2014-rsm}.

When all the bivariate pair-copulas are BVN copulas with correlation (copula) parameters  $\rho_{101,011}$, $\rho_{011,111}$, 
$\rho_{111,100}$,  $\rho_{100,010}$, $\rho_{010,110}$, 
the resulting distribution is the 6-variate normal with mean vector 

\begin{small}
$$
\bmu=\Bigl(l(\pi_{101},\pi_{011},\pi_{111}),l(\pi_{011},\pi_{101},\pi_{111}),l(\pi_{111},\pi_{101},\pi_{111}),
l(\pi_{100},\pi_{010},\pi_{110}),l(\pi_{010},\pi_{100},\pi_{110}),l(\pi_{110},\pi_{100},\pi_{110})
\Bigr)
$$
\end{small}
and variance covariance matrix
\begin{small}
\begin{displaymath}
\Sbf=\begin{pmatrix}
\sigma_{101}^2 &\rho_{101,011}\sigma_{101}\s_{011} &\rho_{101,111}\sigma_{101}\s_{111}&\rho_{101,100}\sigma_{101}\s_{100}&\rho_{101,010}\sigma_{101}\s_{010}&\rho_{101,110}\sigma_{101}\s_{110}\\\\
\rho_{101,011}\sigma_{101}\sigma_{011} & \sigma_{011}^2&\rho_{011,111}\sigma_{011}\s_{111}&\rho_{011,100}\sigma_{011}\s_{100}&\rho_{011,010}\sigma_{011}\s_{010}&\rho_{011,110}\sigma_{011}\s_{110}\\\\
\rho_{101,111}\sigma_{101}\sigma_{111} &\rho_{011,111}\sigma_{011}\s_{111}& \sigma_{111}^2 &\rho_{111,100}\sigma_{111}\sigma_{100}&\rho_{111,010}\sigma_{111}\s_{010}&\rho_{111,110}\sigma_{111}\s_{110}\\\\
\rho_{101,100}\sigma_{101}\sigma_{100}&\rho_{011,100}\sigma_{011}\sigma_{100}  & \rho_{111,100}\sigma_{111}\sigma_{100}
&\sigma_{100}^2&\rho_{100,010}\sigma_{100}\s_{010}&\rho_{100,110}\sigma_{100}\s_{110}\\\\
\rho_{101,010}\sigma_{101}\sigma_{010}&\rho_{011,010}\sigma_{011}\sigma_{010}  & \rho_{111,010}\sigma_{111}\sigma_{010}& \rho_{100,010}\sigma_{100}\sigma_{010}
&\sigma_{010}^2&\rho_{010,110}\sigma_{010}\s_{110}\\\\
\rho_{101,110}\sigma_{101}\sigma_{110}&\rho_{011,110}\sigma_{011}\sigma_{110}  & \rho_{111,110}\sigma_{111}\sigma_{110}& \rho_{100,110}\sigma_{100}\sigma_{110}
&\rho_{010,110}\sigma_{010}\s_{110}&\sigma_{110}^2
\end{pmatrix}, 
\end{displaymath}
\end{small}
where 
\begin{eqnarray*}
\rho_{101,111}=\rho_{101,011}\rho_{011,111} &&\rho_{101,100}=\rho_{101,111}\rho_{111,100}\\ \rho_{101,010}=\rho_{101,100}\rho_{010,110} && \rho_{101,110}=\rho_{101,010}\rho_{010,110}\\  
\rho_{011,100}=\rho_{011,111}\rho_{111,100} &&\rho_{011,010}=\rho_{011,100}\rho_{100,010}\\ 
\rho_{011,110}=\rho_{011,010}\rho_{010,110}&& \rho_{111,010}=\rho_{111,100}\rho_{100,010}\\ \rho_{111,110}=\rho_{111,010}\rho_{010,110}&& \rho_{100,110}=\rho_{100,010}\rho_{010,110}.
\end{eqnarray*}

Covariance and correlation matrices as above play a central role in multivariate Gaussian structures. Nevertheless, two major difficulties in modelling such matrices are multidimensionality, as the number of parameters grows quadratically with dimension, and positive definiteness. 
Our approach  overcomes both difficulties.  Multidimensionality is controlled by focusing on a structured correlation matrix. As we use truncation, a structured correlation matrix  is exploited and thus 5 instead of 15 dependence parameters 
have to be estimated, which is extremely useful as the sample size in our motivating example is so small ($N=11$).  \cite{trikalinos-etal-2014-rsm}, in order to reduce the  parameters even further, 
proposed another structured variant by setting variances and correlations to be equal. 
Furthermore, our parametrization of the $6$-variate Gaussian distribution as a truncated vine consists of algebraically independent correlations 
and 
avoids the positive definite constraints (e.g., \citealt[page 119]{joe2014}).

Hence,  our model has as special case  the  multinomial GLMM with a structured correlation matrix. \cite{trikalinos-etal-2014-rsm} acknowledged that a more direct approach is to model the probabilities on the original scale in the form of a Dirichlet  or multivariate beta distribution and left this for future research. In the preceding section we explicitly develop this method by  using a truncated D-vine copula with beta margins  representation   of the multivariate beta distribution.

\subsection{\label{beta-model}The multinomial truncated  D-vine copula mixed model with beta margins}
Both the multinomial truncated D-vine CMM with normal margins and the multinomial GLMM assume the vector of probabilities for each combination of test results in diseased and non-diseased patients are on a transformed scale. 
However, by using a  copula with beta margins representation of the random effects distribution,  we can model   the latent proportions on their original scale.  As these proportions have unit sum constraints, we  choose to elicit the random effects distribution over the  conditional latent proportions that  have algebraic independence  using the transformation  proposed by 
  \cite{wilson2018}.  

The diseased and non-diseased  subjects  fall into  4 possible categories as indicated in the first two columns  of Table \ref{4times2}. 
Assume that     $\X_{1}$  and $\X_{0}$ is the latent vector of probabilities that the diseased and non-diseased  subjects, respectively, fall into each of these categories  conditional on not falling  into any of the previous categories (rows). We can then recover the original latent proportions via
$$X_{101}\quad X_{011}(1-X_{101}) \quad X_{111}(1-X_{011})(1-X_{101})\quad (1-X_{111})(1-X_{011})(1-X_{101})$$
and
$$X_{100}\quad X_{010}(1-X_{100}) \quad X_{110}(1-X_{010})(1-X_{100})\quad (1-X_{110})(1-X_{010})(1-X_{100}).$$
Clearly two of the latent proportions remain on the original scale, but by permuting $\{10t,01t,11t\}$ we can eventually get all the latent proportions on the original scale.

The within-study model takes the form 
\begin{align}\label{within-diseased-beta}
(Y_{i101},&Y_{i011},
Y_{111},Y_{i001})|(X_{101}=x_{101},X_{011}=x_{011},X_{111}=x_{111})\sim\nonumber\\ &\mathcal{M}_4\Bigl(y_{i++1},x_{101},x_{011}(1-x_{101}),x_{111}(1-x_{011})(1-x_{101})
\Bigr);\\
(Y_{i100},&Y_{i010},
Y_{110},Y_{i000})|(X_{100}=x_{100},X_{010}=x_{010},X_{110}=x_{110})\sim\nonumber\\& \mathcal{M}_4\Bigl(y_{i++0},x_{100},x_{010}(1-x_{100}),x_{110}(1-x_{010})(1-x_{100})
\Bigr)\nonumber
\end{align}

The stochastic representation of the between studies model  is
\begin{align}\label{copula-between-beta}
\biggl(F\Bigl(X_{101};\pi_{101},\g_{101}\Bigr),F\Bigl(X_{011};\frac{\pi_{011}}{1-\pi_{101}},\g_{011}\Bigr),F\Bigl(X_{111};\frac{\pi_{111}}{(1-\frac{\pi_{011}}{1-\pi_{101}})(1-\pi_{101})},\g_{111}\Bigr),&\\F\Bigl(X_{100};\pi_{100},\g_{100}\Bigr),F\Bigl(X_{010};\frac{\pi_{010}}{1-\pi_{100}},\g_{010}\Bigr),F\Bigl(X_{110};\frac{\pi_{110}}{(1-\frac{\pi_{010}}{1-\pi_{100}})(1-\pi_{100})},\g_{110}\Bigr)\biggr)&\sim C_6(\cdot;\thbf),\nonumber
\end{align}
where $C_6(\cdot;\thbf)$ is a 6-dimensional truncated D-vine copula with dependence parameter vector $\thbf$ and $F(\cdot;\pi,\g)$ is the cdf of the Beta($\pi,\g$) distribution with $\pi$ the mean and $\g$ the dispersion parameter. The copula parameter vector $\thbf$ has the dependence parameters of the random effects model and they are separated from the univariate parameters $\pibf_t=(\pi_{10t},\pi_{01t},\pi_{11t})$  and $\gbf_t=(\g_{10t},\g_{01t},\g_{11t})$. As in the preceding subsection, the $\pibf_t$'s  are the  actual parameters of interest as the meta-analytic parameters for the TPR and  FPR are functions of these parameters as shown in (\ref{TPR}) and (\ref{FPR}). 
The univariate parameter vectors $\gbf_t$'s   denote the variabilities of the random effects.

The models in (\ref{within-diseased-beta}) and (\ref{copula-between-beta}) together specify a multinomial truncated D-vine CMM with joint likelihood
\begin{align*}
&L(\pibf_1,\pibf_0,\gbf_1,\gbf_0,\thbf|\y_1,\y_0)
=\\
&\prod_{i=1}^N\int_{[0,1]^6}
g\bigl(y_{i101},y_{i011},
y_{i111};y_{i++1},x_{101},x_{011}(1-x_{101}),x_{111}(1-x_{011})(1-x_{101})\bigr)\times\\&g\bigl(y_{i100},y_{i010},
y_{i110};y_{i++0},x_{100},x_{010}(1-x_{100}),x_{110}(1-x_{010})(1-x_{100})\bigr)
f(\x_1,\x_0;\pibf_1,\pibf_0,\gbf_1,\gbf_0,\thbf)d\x_1d\x_0,
\end{align*}
where $f$ is as in (\ref{vine-density}) where we use beta instead of normal marginal distributions.  Below we transform the integral into an integral over a unit hypercube using the inversion method. Hence the joint likelihood becomes
\begin{align*}
&\prod_{i=1}^N\int_{[0,1]^6}
g\biggl(y_{i101},y_{i011},
y_{i111};y_{i++1},F^{-1}(u_{101};\pi_{101},\g_{101}),F^{-1}\Bigl(u_{011};\frac{\pi_{011}}{1-\pi_{101}},\g_{011}\Bigr)\Bigl(1-F^{-1}(u_{101};\pi_{101},\g_{101})\Bigr),\\&F^{-1}\Bigl(u_{111};\frac{\pi_{111}}{(1-\frac{\pi_{011}}{1-\pi_{101}})(1-\pi_{101})},\g_{111}\Bigr)\Bigl(1-F^{-1}\Bigl(u_{011};\frac{\pi_{011}}{1-\pi_{101}},\g_{011}\Bigr)\Bigr)\Bigl(1-F^{-1}(u_{101};\pi_{101},\g_{101})\Bigr)\biggr)\times\\&
g\biggl(y_{i100},y_{i010},
y_{i110};y_{i++0},F^{-1}(u_{100};\pi_{100},\g_{100}),F^{-1}\Bigl(u_{010};\frac{\pi_{010}}{1-\pi_{100}},\g_{010}\Bigr)\Bigl(1-F^{-1}(u_{100};\pi_{100},\g_{100})\Bigr),\\&F^{-1}\Bigl(u_{110};\frac{\pi_{110}}{(1-\frac{\pi_{010}}{1-\pi_{100}})(1-\pi_{100})},\g_{110}\Bigr)\Bigl(1-F^{-1}\Bigl(u_{010};\frac{\pi_{010}}{1-\pi_{100}},\g_{010}\Bigr)\Bigr)\Bigl(1-F^{-1}(u_{100};\pi_{100},\g_{100})\Bigr)\biggr)\times\\&
c(u_{101},u_{011},u_{111},u_{100},
u_{010},u_{110};\thbf)du_{101}du_{011}du_{111}du_{100}du_{010}du_{110}.
\end{align*}

\subsection{\label{computation}Maximum likelihood estimation and computational details}

Estimation of the model parameters    can be approached by the standard maximum likelihood (ML) method, by maximizing the logarithm of the joint likelihood. 
The estimated parameters can be obtained by 
using a quasi-Newton \citep{nash90} method applied to the logarithm of the joint likelihood.   
We use a quasi-Newton method  from the desire to use something like Newton's method for its speed but without having to compute the Hessian matrix each time.  Hence, the quasi-Newton minimization with an input function the negative log-likelihood to be minimized, has  output point of minimum and inverse Hessian at point of minimum.

For the multinomial truncated D-vine CMM numerical evaluation of the joint pmf can be achieved with the following steps:

\begin{enumerate}
\itemsep=0pt
\item Calculate Gauss-Legendre \citep{Stroud&Secrest1966}  quadrature points $\{u_q: q=1,\ldots,N_q\}$ 
and weights $\{w_q: q=1,\ldots,N_q\}$ in terms of standard uniform.
\item Convert from independent uniform random variables $\{u_{q_{101}}: q_{101}=1,\ldots,N_q\}$,  $\{u_{q_2}: q_{011}=1,\ldots,N_q\}$, $\{u_{q_{111}}: q_{111}=1,\ldots,N_q\}$,  $\{u_{q_{100}}: q_{100}=1,\ldots,N_q\}$, $\{u_{q_{010}}: q_{010}=1,\ldots,N_q\}$, and $\{u_{q_{110}}: q_{110}=1,\ldots,N_q\}$ to   dependent uniform random variables $v_{q_{101}},v_{q_{011}|q_{101}},v_{q_{111}|q_{011};q_{101}}$, $v_{q_{100}|q_{111};q_{011},q_{101}}$, $v_{q_{010}|q_{100};q_{111},q_{011},q_{101}}$, and  $v_{q_{110}|q_{010};q_{100},q_{111},q_{011},q_{101}}$ that have a truncated D-vine distribution $C(\cdot;\thbf)$:  
\begin{algorithmic}[1]
\State Set  $v_{q_{101}}=u_{q_{101}}$
\State $v_{q_{011}|q_{101}}=C^{-1}_{{011}|{101}}(u_{q_{011}}|u_{q_{101}};\th_{{101},{011}})$

\State $v_{q_{111}|q_{011};q_{101}}=C_{{111}|{011}}^{-1}(u_{q_{111}}|v_{q_{011}|q_{101}};\th_{{011},{111}})$

\State $v_{q_{100}|q_{111};q_{011},q_{101}}=C^{-1}_{{100}|{111}}(u_{q_{100}}|v_{q_{111}|q_{011};q_{101}};\th_{{111},{100}})$
\State $v_{q_{010}|q_{100};q_{111},q_{011},q_{101}}=C^{-1}_{{010}|{100}}(u_{q_{010}}|v_{q_{100}|q_{111};q_{011},q_{101}};\th_{{100},{010}})$
\State $v_{q_{110}|q_{010};q_{100},q_{111},q_{011},q_{101}}=C^{-1}_{{110}|{010}}(u_{q_{110}}|v_{q_{010}|q_{100};q_{111},q_{011},q_{101}};\th_{{010},{110}})$,
\end{algorithmic}
where   $C^{-1}(v|u;\th)$ are inverse  conditional copula cdfs. The method is  based on the simulation algorithm of a D-vine copula (e.g. \citealt[page 292]{joe2014}), where as input, instead of independent uniform variates, it uses the independent quadrature points.

\item Numerically evaluate the joint pmf, e.g.,
\begin{align*}
&\int_{[0,1]^6}
g\biggl(y_{i101},y_{i011},
y_{i111};y_{i++1},F^{-1}(u_{101};\pi_{101},\g_{101}),F^{-1}\Bigl(u_{011};\frac{\pi_{011}}{1-\pi_{101}},\g_{011}\Bigr)\Bigl(1-F^{-1}(u_{101};\pi_{101},\g_{101})\Bigr),\\&F^{-1}\Bigl(u_{111};\frac{\pi_{111}}{(1-\frac{\pi_{011}}{1-\pi_{101}})(1-\pi_{101})},\g_{111}\Bigr)\Bigl(1-F^{-1}\Bigl(u_{011};\frac{\pi_{011}}{1-\pi_{101}},\g_{011}\Bigr)\Bigr)\Bigl(1-F^{-1}(u_{101};\pi_{101},\g_{101})\Bigr)\biggr)\times\\&
g\biggl(y_{i100},y_{i010},
y_{i110};y_{i++0},F^{-1}(u_{100};\pi_{100},\g_{100}),F^{-1}\Bigl(u_{010};\frac{\pi_{010}}{1-\pi_{100}},\g_{010}\Bigr)\Bigl(1-F^{-1}(u_{100};\pi_{100},\g_{100})\Bigr),\\&F^{-1}\Bigl(u_{110};\frac{\pi_{110}}{(1-\frac{\pi_{010}}{1-\pi_{100}})(1-\pi_{100})},\g_{110}\Bigr)\Bigl(1-F^{-1}\Bigl(u_{010};\frac{\pi_{010}}{1-\pi_{100}},\g_{010}\Bigr)\Bigr)\Bigl(1-F^{-1}(u_{100};\pi_{100},\g_{100})\Bigr)\biggr)\times\\&
c(u_{101},u_{011},u_{111},u_{100},
u_{010},u_{110};\thbf)du_{101}du_{011}du_{111}du_{100}du_{010}du_{110}
\end{align*}

\noindent in a sextuple sum:
\begin{align*}
&\sum_{q_{101}=1}^{N_q}\sum_{q_{011}=1}^{N_q}\sum_{q_{111}=1}^{N_q}
\sum_{q_{100}=1}^{N_q}\sum_{q_{010}=1}^{N_q}\sum_{q_{110}=1}^{N_q}w_{q_{101}}w_{q_{011}}w_{q_{111}}w_{q_{100}}w_{q_{010}}w_{q_{110}}\times
g\biggl(y_{i101},y_{i011},
y_{i111};y_{i++1},\\&F^{-1}(v_{q_{101}};\pi_{101},\g_{101}),F^{-1}\Bigl(v_{q_{011}|q_{101}};\frac{\pi_{011}}{1-\pi_{101}},\g_{011}\Bigr)\Bigl(1-F^{-1}(v_{q_{101}};\pi_{101},\g_{101})\Bigr),F^{-1}\Bigl(v_{q_{111}|q_{011};q_{101}};\\&\frac{\pi_{111}}{(1-\frac{\pi_{011}}{1-\pi_{101}})(1-\pi_{101})},\g_{111}\Bigr)\Bigl(1-F^{-1}\Bigl(v_{q_{011}|q_{101}};\frac{\pi_{011}}{1-\pi_{101}},\g_{011}\Bigr)\Bigr)\Bigl(1-F^{-1}(v_{q_{101}};\pi_{101},\g_{101})\Bigr)\biggr)\times\\&
g\biggl(y_{i100},y_{i010},
y_{i110};y_{i++0},F^{-1}(v_{q_{100}|q_{111};q_{011},q_{101}};\pi_{100},\g_{100}),F^{-1}\Bigl(v_{q_{010}|q_{100};q_{111},q_{011},q_{101}};\frac{\pi_{010}}{1-\pi_{100}},\g_{010}\Bigr)\times\\&\Bigl(1-F^{-1}(v_{q_{100}|q_{111};q_{011},q_{101}};\pi_{100},\g_{100})\Bigr),F^{-1}\Bigl(v_{q_{110}|q_{010};q_{100},q_{111},q_{011},q_{101}};\frac{\pi_{110}}{(1-\frac{\pi_{010}}{1-\pi_{100}})(1-\pi_{100})},\g_{110}\Bigr)\times\\&\Bigl(1-F^{-1}\Bigl(v_{q_{010}|q_{100};q_{111},q_{011},q_{101}};\frac{\pi_{010}}{1-\pi_{100}},\g_{010}\Bigr)\Bigr)\Bigl(1-F^{-1}(v_{q_{100}|q_{111};q_{011},q_{101}};\pi_{100},\g_{100})\Bigr)\biggr)
\end{align*}

\end{enumerate}

With Gauss-Legendre quadrature, the same nodes and weights
are used for different functions;
this helps in yielding smooth numerical derivatives for numerical optimization via quasi-Newton.

\section{\label{simulations-section}Small-sample efficiency -- misspecification of the random effects distribution}
In this section a simulation study with two different scenarios is conducted to (a) assess the performance of  the ML method in Section \ref{computation}, and (b) investigate  the effect of the misspecification of the parametric margin of the random effects distribution. 

We set the sample size and the  true univariate  and dependence parameters to mimic the motivating  
example analysed in Section \ref{application} and  use the following simulation process:

\begin{enumerate}
\itemsep=10pt
 \item Simulate $(u_{101},u_{011},u_{111},u_{100},
u_{010},u_{110})$ from a truncated D-vine distribution $C(\cdot;\thbf)$. 
\item 

\begin{itemize}
\itemsep=10pt
\item Convert to  normal realizations via 
\begin{align*}
&x_{101}=\Phi^{-1}\Bigl(u_{101};l(\pi_{101},\pi_{011},\pi_{111}),\s_{101}^2\Bigr)\quad x_{011}=\Phi^{-1}\Bigl(u_{011};l(\pi_{011},\pi_{101},\pi_{111}),\s_{011}^2\Bigr)\\&x_{111}=\Phi^{-1}\Bigl(u_{111};l(\pi_{111},\pi_{101},\pi_{011}),\s_{111}^2\Bigr)\quad x_{100}=\Phi^{-1}\Bigl(u_{100};l(\pi_{100},\pi_{010},\pi_{110}),\s_{100}^2\Bigr)\\&x_{010}=\Phi^{-1}\Bigl(u_{010},l(\pi_{010},\pi_{100},\pi_{110}),\s_{010}^2\Bigr)\quad x_{110}=\Phi^{-1}\Bigl(u_{110};l(\pi_{110},\pi_{100},\pi_{010}),\s_{110}^2\Bigr).
\end{align*}

\item Convert to  beta realizations via 
\begin{align*}
&x_{101}=F^{-1}\Bigl(u_{101};\pi_{101},\g_{101}\Bigr)\qquad  x_{011}=F^{-1}\Bigl(u_{011};\frac{\pi_{011}}{1-\pi_{101}},\g_{011}\Bigr)\\&x_{111}=F^{-1}\Bigl(u_{111};\frac{\pi_{111}}{(1-\frac{\pi_{011}}{1-\pi_{101}})(1-\pi_{101})},\g_{111}\Bigr)\qquad x_{100}=F^{-1}\Bigl(u_{100};\pi_{100},\g_{100}\Bigr)\\&x_{010}=F^{-1}\Bigl(u_{010};\frac{\pi_{010}}{1-\pi_{100}},\g_{010}\Bigr)\qquad x_{110}=F^{-1}\Bigl(u_{110};\frac{\pi_{110}}{(1-\frac{\pi_{010}}{1-\pi_{100}})(1-\pi_{100})},\g_{110}\Bigr)
\end{align*}

\end{itemize} 
\item Simulate the  size of diseased and non-diseased subjects  $n_1$  and $n_0$, respectively, from a shifted gamma distribution  to obtain heterogeneous study sizes \citep{paul-etal-2010},  i.e., 
\begin{eqnarray*}
n_1&\sim& \mbox{sGamma}(\a=1.2,\b=0.01,\mbox{lag}=30)\\
n_0&\sim& \mbox{sGamma}(\a=1.2,\b=0.01,\mbox{lag}=30)
\end{eqnarray*}
and round off  $n_1$  and $n_0$ to the nearest integers. 
\item 
\begin{itemize}
\itemsep=20pt
\item For normal margins draw $(y_{101},y_{011},
y_{111},y_{001})$ from
$$\mathcal{M}_4\Bigl(n_1,
l^{-1}(x_{101},x_{011},x_{111}),
l^{-1}(x_{011},x_{101},x_{111}),
l^{-1}(x_{111},x_{101},x_{011})\Bigr)$$

and  $(y_{100},y_{010},y_{110},
y_{000})$ from 
$$\mathcal{M}_4\Bigl(n_0,
l^{-1}(x_{100},x_{010},x_{110}),
l^{-1}(x_{010},x_{100},x_{110}),
l^{-1}(x_{110},x_{100},x_{010})\Bigr)$$

\item For beta margins draw $(y_{101},y_{011},
y_{111},y_{001})$ from  
$$\mathcal{M}_4\Bigl(n_1,x_{101},x_{011}(1-x_{101}),x_{111}(1-x_{011})(1-x_{101})\Bigr)$$

 and  $(y_{100},y_{010},y_{110},
y_{000})$  from 
$$\mathcal{M}_4\Bigl(n_0,x_{100},x_{010}(1-x_{100}),x_{110}(1-x_{010})(1-x_{100})\Bigr)$$

\end{itemize}

\end{enumerate}

In the first scenario the simulated data are generated from a multinomial truncated D-vine CMM with BVN copulas and normal margins (the resulting model is the same with the multinomial GLMM in  \citealt{trikalinos-etal-2014-rsm}), while in the second scenario the simulated data are generated  from a multinomial truncated D-vine CMM with BVN copulas and beta margins. 
Table  \ref{sim}  contains the 
resultant biases, root mean square errors (RMSE), and standard deviations (SD),  along with the square root of the average theoretical variances ($\sqrt{\bar V}$), scaled by 100, for the ML estimates  of  the multinomial truncated D-vine CMM with BVN copulas and normal margins, i.e., the multinomial GLMM.  The theoretical variances of the MLEs for each simulated dataset is obtained via the gradients and the Hessian computed numerically during the quasi-Newton minimization.  Note that we converted from the BVN copula parameters 
$\theta$'s to $\tau$'s  via the relation 
\begin{equation}\label{tauBVN}
\tau=\frac{2}{\pi}\arcsin(\th).
\end{equation}

We have simulated from normal margins and estimated with normals margins (left side of Table 2) or have simulated from beta margins  and estimated with normals  margins (right side of the Table 2). 
A complete cross-over simulation, where we have also simulated from either beta or normal margins and estimated with beta margins is provided in \cite{Nikoloulopoulos-2018-4dmeta-NE}  for  the quadrivariate multinomial D-vine CMM. Same results are to be expected for the 6-dimensional case.

\begin{table}[!h]
  \centering
  \caption{\label{sim} Small sample of sizes $N = 11$ simulations ($10^3$ replications, $N_q=15$) from the multinomial truncated D-vine copula mixed  model   with BVN copulas and both  normal (that is the same with the multinomial GLMM) and beta margins and biases,  root mean square errors (RMSEs) and standard deviations (SDs), along with the square root of the average theoretical variances ($\sqrt{\bar V}$), scaled by 100, for the MLEs of the multinomial GLMM.  }

 \setlength{\tabcolsep}{5pt}  
  
\begin{small}
\begin{tabular}{ccccccccccc}
   \toprule
   \multicolumn{11}{c}{True (simulated) univariate margin} \\
   \midrule
   \multicolumn{5}{c}{normal}            &       & \multicolumn{5}{c}{beta} \\\cmidrule{1-5} \cmidrule{7-11}
 True values        & Bias & SD    & $\sqrt{\bar V}$ & RMSE  &       &    True values   & Bias & SD    & $\sqrt{\bar V}$ & RMSE \\\hline
 $\pi_{101}=0.037$  & 0.103 & 4.184 & 1.611 & 4.185 &       &  $\pi_{101}=0.091$  & -6.181 & 2.787 & 1.529 & 6.781 \\
       $\pi_{011}=0.093$  & 0.449 & 2.125 & 1.929 & 2.172 &       &  $\pi_{011}=0.086$  & 0.773 & 2.154 & 2.164 & 2.289 \\
      $\pi_{111}=0.295$  & 0.671 & 4.769 & 3.685 & 4.816 &       &  $\pi_{111}=0.292$  & 1.716 & 4.445 & 3.615 & 4.765 \\
       $\pi_{100}=0.017$  & 0.128 & 0.753 & 0.673 & 0.764 &       &  $\pi_{100}=0.024$ & -0.461 & 0.724 & 0.616 & 0.858 \\
       $\pi_{010}=0.049$  & 0.067 & 0.995 & 0.915 & 0.997 &       &  $\pi_{010}=0.054$ & -0.357 & 1.014 & 0.874 & 1.075 \\
     $\pi_{110}=0.030$  & 0.026 & 0.744 & 0.653 & 0.745 &       &  $\pi_{110}=0.034$  & -0.336 & 0.724 & 0.637 & 0.798 \\
     $\pi_{1\cdot1}=0.331$  & 0.774 & 4.232 & 3.364 & 4.303 &       & $\pi_{1\cdot1}=0.383$  & -4.465 & 4.204 & 3.444 & 6.133 \\
      $\pi_{\cdot11}=0.388$  & 1.121 & 5.898 & 4.422 & 6.003 &       & $\pi_{\cdot11}=0.378$  & 2.489 & 5.476 & 4.444 & 6.015 \\
       $\pi_{1\cdot0}=0.047$  & 0.154 & 1.064 & 0.927 & 1.075 &       & $\pi_{1\cdot0}=0.058$  & -0.797 & 1.014 & 0.881 & 1.290 \\
     $\pi_{\cdot10}=0.079$  & 0.093 & 1.443 & 1.204 & 1.446 &       & $\pi_{\cdot10}=0.088$  & -0.693 & 1.430 & 1.157 & 1.589 \\
       $\s_{101}=1.699$  & -16.076 & 59.052 & 30.366 & 61.201 &       & $\g_{101}=0.186$  & -     & 60.137 & 38.151 & - \\
       $\s_{011}=0.543$  & 13.386 & 36.956 & 21.917 & 39.306 &       & $\g_{011}=0.016$  & -     & 36.311 & 27.041 & - \\
       $\s_{111}=0.585$  & -3.872 & 23.458 & 15.216 & 23.775 &       & $\g_{111}=0.066$  & -     & 20.195 & 15.013 & - \\
       $\s_{100}=0.929$  & -9.645 & 37.941 & 37.035 & 39.148 &       & $\g_{100}=0.015$  & -     & 35.120 & 35.450 & - \\
       $\s_{010}=0.490$  & -6.043 & 23.022 & 21.888 & 23.802 &       & $\g_{010}=0.011$  & -     & 22.394 & 20.545 & - \\
       $\s_{110}=0.570$  & -8.459 & 27.213 & 22.832 & 28.497 &       & $\g_{110}=0.010$  & -     & 26.296 & 23.279 & - \\
       $\tau_{101,011}=-0.525$  & 5.066 & 31.093 & 32.432 & 31.503 &       & $\tau_{101,011}=-0.525$  & 16.168 & 31.319 & 29.155 & 35.246 \\
       $\tau_{011,111}=0.558$  & -5.571 & 31.463 & 27.902 & 31.952 &       & $\tau_{011,111}=0.300$  & 15.472 & 29.684 & 29.010 & 33.474 \\
       $\tau_{111,100}=0.185$  & 0.800 & 37.174 & 36.929 & 37.183 &       & $\tau_{111,100}=0.197$  & 2.185 & 39.347 & 38.178 & 39.408 \\
       $\tau_{100,010}=0.022$  & 4.256 & 41.838 & 55.494 & 42.053 &       & $\tau_{100,010}=-0.029$  & 7.018 & 44.426 & 56.755 & 44.977 \\
       $\tau_{010,110}=0.576$  & -7.875 & 42.820 & 72.731 & 43.538 &       & $\tau_{010,110}=0.544$  & -2.784 & 40.708 & 64.822 & 40.803 \\
   \bottomrule
   \end{tabular}%
   \end{small}
 \label{tab:addlabel}%
\end{table}

Conclusions from the values in the table are the following:
\begin{itemize}
\itemsep=0pt
\item  ML with the true multinomial truncated D-vine CMM is highly efficient 
according to the simulated biases, SDs and RMSEs. For example in Table \ref{sim} where the true univariate margins are normal the scaled biases for the MLEs of $\pibf_{0}$ for the multinomial truncated  vine CMM with  BVN copulas and  normal margins range from $0.026$ to $0.128$. 
\item The ML estimates of 
the univariate parameters of the main interest $\pibf_1,\pibf_0$ and their functions   $\pi_{1\cdot t},  \pi_{\cdot1t}$ 
are not robust to margin misspecification, e.g., in Table 2 where the true univariate margins are beta the scaled biases for the MLEs of $\pibf_{1}$ for the multinomial truncated  vine CMM with  BVN copulas and  normal margins range from $-6.181$ to $1.716$.

\item The ML estimates of $\tau$'s  
are not robust to margin misspecification, e.g., in Table 2 where the true univariate margins are beta the scaled biases for the MLEs of $\tau$'s for the multinomial truncated  vine CMM with  BVN copulas and  normal margins range from $-2.784$ to $16.168$.

\item The SDs and $\sqrt{\bar V}$'s   for the Kendall's $\tau$ and variability parameters are larger. 
 This is due to the typical small sample size $N=11$ for estimating 6 variance and 5  Kendall's $\tau$ parameters on the top of the 6 probability parameters that are of the main interest.  
\cite{trikalinos-etal-2014-rsm} also acknowledged these parameters are often not well estimated.

\end{itemize}

The simulation results indicate that  the effect of misspecifying the marginal choice can be seen as substantial for both the univariate parameters of interest and their functions in (\ref{TPR}) and (\ref{FPR}). Hence, the multinomial GLMM can lead to biased meta-analytic estimates of interest and their functions as it is restricted to a normal margin specification. This is in line with   our previous studies in CMMs \citep{Nikoloulopoulos2015b,Nikoloulopoulos2015c,Nikoloulopoulos-2016-SMMR,Nikoloulopoulos2018-AStA,nikoloulopoulos-2018-smmr,Nikoloulopoulos-2018-3dmeta-NE,Nikoloulopoulos-2018-4dmeta-NE,Nikoloulopoulos2020-factorREMADA}.

 \cite{Nikoloulopoulos-2018-4dmeta-NE} performed extensive simulations to study  the misspecification of the parametric bivariate copula of the random effects distribution of the  quadrivariate multinomial D-vine CMM. Same results are to be expected for the 6-dimensional case. It has been shown that the effect of misspecifying the copula choice can be seen as minimal for both the univariate parameters and Kendall's tau, which  is a strictly increasing function of the copula parameter for any pair-copula,  as  (a) the meta-analytic parameters are a univariate inference, and hence, it is the univariate marginal distribution that matters and not the type of the pair-copula, and (b) Kendall’s tau only accounts for the dependence dominated by the middle of the data, and it is expected to be similar amongst different families of bivariate copulas. However, the tail dependence varies, and is a property to consider when choosing amongst different families of bivariate copulas. Any inference that depends on the joint distribution  will essentially show the effect of different model (random effect distribution) assumptions such as the pair-copula choice. We discuss 
such an inference in the forthcoming section. 

\section{\label{SROC-section}Summary receiver operating characteristic curves}
Though typically the focus of meta-analysis has been to derive the  summary-effect estimates, there is increasing interest in alternative summary outputs, such as summary receiver operating characteristic (SROC) curves.  \cite{trikalinos-etal-2014-rsm} haven't derived the 
SROC curves  from the  the multinomial GLMM 
as the latent vector of  probabilities of each combination of test results in diseased and non-diseased patients is on a transformed scale via the multinomial logit link.  

In this section we derive the SROC curves  from the multinomial truncated D-vine CMM with beta margins taking advantage of the fact that 
some of the latent proportions can be on a the original scale. 
We have to first strategically permute the variables as $X_{11t},X_{10t},X_{01t}$, so that $X_{111}$ and $X_{110}$ are on the original scale. Hence, the  
within-study  and between studies  models   take the form  
\begin{align*}
(Y_{111},Y_{i101},&Y_{i011},
Y_{i001})|(X_{111}=x_{111},X_{101}=x_{101},X_{011}=x_{011})\sim\nonumber\\ &\mathcal{M}_4\Bigl(y_{i++1},x_{111},x_{101}(1-x_{111}),x_{011}(1-x_{101})(1-x_{111})
\Bigr);\\
(Y_{110},Y_{i100},&Y_{i010},
Y_{i000})|(X_{110}=x_{110},X_{100}=x_{100},X_{010}=x_{010})\sim\nonumber\\& \mathcal{M}_4\Bigl(y_{i++0},x_{110},x_{100}(1-x_{110}),x_{010}(1-x_{010})(1-x_{110})
\Bigr)\nonumber
\end{align*}
and 
\begin{align*}
&\biggl(F\Bigl(X_{111};\pi_{111},\g_{111}\Bigr),F\Bigl(X_{110};\pi_{110},\g_{110}\Bigr),F\Bigl(X_{101};\frac{\pi_{101}}{1-\pi_{111}},\g_{101}\Bigr),\\&F\Bigl(X_{011};\frac{\pi_{011}}{(1-\frac{\pi_{101}}{1-\pi_{111}})(1-\pi_{111})},\g_{011}\Bigr)F\Bigl(X_{100};\frac{\pi_{100}}{1-\pi_{110}},\g_{100}\Bigr),\\&F\Bigl(X_{010};\frac{\pi_{010}}{(1-\frac{\pi_{100}}{1-\pi_{110}})(1-\pi_{110})},\g_{010}\Bigr)\biggr)\sim C_6(\cdot;\th_{111,110},\th_{110,101},\th_{101,011},\th_{011,100},\th_{100,010}),\nonumber
\end{align*}
respectively. With this permutation we achieve  that $X_{11t}\sim B(\pi_{11t},\g_{11t})$ and  the bivariate copula that  links  $X_{111}$ and $X_{110}$ is $C_2\Bigl(F(X_{111};\pi_{111},\g_{111}),F(X_{110};\pi_{110},\g_{110});\theta_{111,110}\Bigr)$.

Then, the  SROC
 curves for the  latent pair $(X_{111}, X_{110})$  can be deduced 
 through the quantile regression techniques developed in \cite{Nikoloulopoulos2015b}:

 \begin{enumerate}
\item Set $C_2(u_{111}|u_{110};\th_{111,110})=q$;
\item Solve for the quantile regression curve $u_{111}:=\widetilde{u}_{111}(u_{110},q;\th_{111,110})=C^{-1}(q|u_{110};\th_{111,110})$;
\item Replace $u_{11t}$ by $F\Bigl(x_{11t};\pi_{11t},\g_{11t}\Bigr)$; 
\item Plot $x_{111}:=\widetilde{x}_{111}(x_{110},q)$ versus $x_{110}$. 
\end{enumerate}

As there is no priori reason to regress $X_{111}$ on $X_{110}$ instead of the other way around \citep{Arends-etal-2008},  quantile regression   curves of $X_{110}$ on $X_{111}$ are also derived in a similar manner.  We use the median regression curves  ($q=0.5$), along with the quantile regression curves with a focus on high ($q$ = 0.99) and low quantiles ($q$ = 0.01), which are strongly associated with the upper and lower tail dependence, respectively, imposed from each parametric family of bivariate copulas. These can be seen as confidence regions, as per the terminology  in \cite{Rucker-schumacher-2009},   of  the median regression curves. 
Finally, in order  to reserve the nature of a bivariate response instead of a univariate response along with a covariate, we plot the corresponding  contour graph of the bivariate copula density with beta margins. The contour plot can be seen as the predictive region (analogously  to \citealt{Reitsma-etal-2005}) of the estimated pair  $(\pi_{111},\pi_{110})$ of the meta-analytic parameters of joint TPR  and joint FPR.

\section{\label{application}Joint meta-analysis of shortened humerus  and shortened femur  of the fetus markers}
We demonstrate the multinomial truncated D-vine CMM by insightfully re-analysing the data on $N=11$ studies from a systematic review that examined  the screening accuracy of shortened humerus  and shortened femur  of the fetus markers (two out of seven ultrasonographic markers or their combination in detecting Down syndrome in  \citealt{Smith-Bindman-etal-2001}). These data have been previously used  to motivate the multinomial GLMM for the joint meta-analysis of multiple tests and are available in the supplementary material of \cite{trikalinos-etal-2014-rsm}.

We fit the multinomial truncated D-vine CMM for all different 360  decompositions of the truncated D-vine copula density, for both beta and normal margins and different pair copulas. In our  general statistical model 
there are no constraints in the choices of the parametric marginal  or pair-copula  distributions.
This is one of the limitations of the multinomial GLMM where all the pair copulas are BVN and marginal distributions are normal.  For ease of interpretation, we do not mix pair-copulas or margins and in line with our previous contributions in CMMs \citep{Nikoloulopoulos2015b,Nikoloulopoulos2015c,Nikoloulopoulos-2016-SMMR,Nikoloulopoulos2018-AStA,nikoloulopoulos-2018-smmr,Nikoloulopoulos-2018-3dmeta-NE,Nikoloulopoulos-2018-4dmeta-NE,Nikoloulopoulos2020-factorREMADA} we use 
 bivariate parametric copulas with different tail dependence behaviour, namely the BVN with intermediate tail dependence, Frank with tail independence, and Clayton with positive lower tail dependence. For the latter we also use its rotated versions to provide negative upper-lower tail dependence (Clayton rotated by 90$^\circ$), positive upper tail dependence (Clayton rotated by 180$^\circ$) and negative lower-upper tail dependence (Clayton rotated by 270$^\circ$). To make it easier
to compare strengths of dependence amongst different copulas, we convert from the BVN
$\theta$'s to $\tau$'s  via the relation in (\ref{tauBVN}) and from the Frank  and   (rotated)  Clayton $\theta$'s to $\tau$'s  via the relations
$$
\tau=\left\{\begin{array}{ccc}
1-4\theta^{-1}-4\theta^{-2}\int_\theta^0\frac{t}{e^t-1}dt &,& \th<0\\
1-4\theta^{-1}+4\theta^{-2}\int^\theta_0\frac{t}{e^t-1}dt &,& \th>0\\
\end{array}\right.
\quad 
\mbox{and} \quad  
\tau=\left\{\begin{array}{rcl}
\th/(\th+2)&,& \mbox{by 0$^\circ$ or 180$^\circ$ }\\
-\th/(\th+2)&,& \mbox{by 90$^\circ$ or 270$^\circ$}\\
\end{array}\right.
$$
in  \cite{genest87} and 
\cite{genest&mackay86}, respectively. 
 In cases when fitting the multinomial truncated D-vine CMM, the resultant estimate of one of the  Kendall's $\tau$ parameters was close to the right ($0.95$) or left boundary ($-0.95$) of its parameter space, we set the corresponding bivariate copula to comonotonic (Fr\'echet lower bound) or countermonotonic  (Fr\'echet lower bound) copula, respectively,  and, then optimize over the remaining  parameters.

\begin{table}[!h]
  \centering
  \caption{\label{app-table}Maximized log-likelihoods, estimates and standard errors (SE), of the multinomial truncated D-vine copula mixed models for the accuracy of  shortened humerus  and shortened femur  of the fetus  in detecting Down syndrome in liveborn infants.}

  \setlength{\tabcolsep}{2.6pt}
\begin{small}
    \begin{tabular}{ccccccccccccccccccc}
    \toprule
          &       & \multicolumn{2}{c}{BVN} &       & \multicolumn{2}{c}{Frank} &       & \multicolumn{2}{c}{Cln\{$180^\circ,90^\circ$\}} &       & \multicolumn{2}{c}{Cln\{$180^\circ,270^\circ$\}} &       & \multicolumn{2}{c}{Cln\{$0^\circ,90^\circ$\}} &       & \multicolumn{2}{c}{Cln\{$0^\circ,270^\circ$\}} \\
    \cmidrule{3-4}  \cmidrule{6-7} \cmidrule{9-10}
    \cmidrule{12-13} \cmidrule{15-16} \cmidrule{18-19}
          &       & Est.  & SE    &       & Est.  & SE    &       & Est.  & SE    &       & Est.  & SE    &       & Est.  & SE    &       & {Est.} & {SE} \\\midrule
    normal margins &       &       &       &       &       &       &       &       &       &       &       &       &       &       &       &       & {} & {} \\
    $\pi_{101}$ &       & 0.037 & 0.026 &       & 0.034 & 0.024 &       & 0.036 & 0.025 &       & 0.037 & 0.025 &       & 0.040 & 0.024 &       & {0.036} & {0.026} \\
    $\pi_{011}$ &       & 0.093 & 0.026 &       & 0.094 & 0.024 &       & 0.083 & 0.022 &       & 0.090 & 0.025 &       & 0.080 & 0.022 &       & {0.099} & {0.019} \\
   $\pi_{111}$ &       & 0.295 & 0.048 &       & 0.299 & 0.047 &       & 0.298 & 0.048 &       & 0.295 & 0.047 &       & 0.312 & 0.041 &       & {0.308} & {0.042} \\
    $\pi_{100}$ &       & 0.017 & 0.006 &       & 0.017 & 0.006 &       & 0.017 & 0.005 &       & 0.017 & 0.005 &       & 0.017 & 0.005 &       & {0.017} & {0.006} \\
    $\pi_{010}$ &       & 0.049 & 0.007 &       & 0.046 & 0.006 &       & 0.047 & 0.008 &       & 0.047 & 0.008 &       & 0.047 & 0.006 &       & {0.047} & {0.006} \\
    $\pi_{110}$ &       & 0.030 & 0.006 &       & 0.029 & 0.005 &       & 0.029 & 0.005 &       & 0.029 & 0.005 &       & 0.030 & 0.005 &       & {0.030} & {0.005} \\
      $\pi_{1\cdot1}$ &       & 0.331 & 0.042 &       & 0.333 & 0.041 &       & 0.334 & 0.041 &       & 0.332 & 0.039 &       & 0.351 & 0.044 &       & {0.345} & {0.041} \\
  $\pi_{\cdot11}$ &       & 0.388 & 0.057 &       & 0.393 & 0.055 &       & 0.381 & 0.056 &       & 0.385 & 0.055 &       & 0.392 & 0.042 &       & {0.407} & {0.046} \\
   $\pi_{1\cdot0}$ &       & 0.047 & 0.008 &       & 0.046 & 0.007 &       & 0.046 & 0.007 &       & 0.046 & 0.007 &       & 0.048 & 0.008 &       & {0.048} & {0.008} \\
 $\pi_{\cdot10}$ &       & 0.079 & 0.011 &       & 0.075 & 0.009 &       & 0.076 & 0.011 &       & 0.076 & 0.011 &       & 0.077 & 0.010 &       & {0.077} & {0.010} \\
    $\s_{101}$ &       & 1.699 & 0.670 &       & 1.670 & 0.688 &       & 1.653 & 0.654 &       & 1.648 & 0.662 &       & 1.524 & 0.610 &       & {1.708} & {0.679} \\
    $\s_{011}$ &       & 0.543 & 0.303 &       & 0.522 & 0.287 &       & 0.678 & 0.282 &       & 0.567 & 0.305 &       & 0.530 & 0.276 &       & {0.368} & {0.188} \\
    $\s_{111}$ &       & 0.585 & 0.189 &       & 0.578 & 0.186 &       & 0.584 & 0.180 &       & 0.571 & 0.172 &       & 0.460 & 0.135 &       & {0.492} & {0.145} \\
    $\s_{100}$ &       & 0.929 & 0.266 &       & 0.939 & 0.243 &       & 0.925 & 0.234 &       & 0.926 & 0.234 &       & 0.913 & 0.256 &       & {0.926} & {0.260} \\
    $\s_{010}$ &       & 0.490 & 0.104 &       & 0.411 & 0.075 &       & 0.504 & 0.144 &       & 0.504 & 0.131 &       & 0.400 & 0.067 &       & {0.398} & {0.067} \\
    $\s_{110}$ &       & 0.570 & 0.160 &       & 0.521 & 0.144 &       & 0.549 & 0.157 &       & 0.547 & 0.155 &       & 0.531 & 0.150 &       & {0.532} & {0.150} \\
    $\tau_{101,011}$ &       & -0.525 & 0.480 &       & -0.531 & 0.440 &       & -0.756 & 0.457 &       & -0.647 & 0.494 &       & -0.724 & 0.679 &       & {-0.420} & {0.332} \\
    $\tau_{011,111}$ &       & 0.558 & 0.458 &       & 0.572 & 0.458 &       & 0.479 & 0.269 &       & 0.548 & 0.392 &       & 0.000 & 0.167 &       & {0.95} & {-} \\
    $\tau_{111,100}$ &       & 0.185 & 0.285 &       & 0.097 & 0.266 &       & 0.023 & 0.244 &       & 0.029 & 0.258 &       & 0.321 & 0.120 &       & {0.281} & {0.226} \\
    $\tau_{100,010}$ &       & 0.022 & 0.201 &       & 0.113 & 0.211 &       & 0.000 & 0.285 &       & 0.000 & 0.235 &       & 0.134 & 0.156 &       & {0.134} & {0.162} \\
    $\tau_{010,110}$ &       & 0.576 & 0.178 &       & 0.629 & 0.185 &       & 0.427 & 0.259 &       & 0.426 & 0.238 &       & 0.705 & 0.219 &       & {0.689} & {0.226} \\
 
    $-\log L$ &       & \multicolumn{2}{c}{3192.90} &       & \multicolumn{2}{c}{3192.57} &       & \multicolumn{2}{c}{3193.73} &       & \multicolumn{2}{c}{3193.76} &       & \multicolumn{2}{c}{3192.44} &       & \multicolumn{2}{c}{3191.94} \\
    beta margins &       &       &       &       &       &       &       &       &       &       &       &       &       &       &       &       & {} & {} \\
    $\pi_{101}$ &       & 0.091 & 0.044 &       & 0.088 & 0.042 &       & 0.086 & 0.040 &       & 0.090 & 0.043 &       & 0.091 & 0.043 &       & 0.093 & 0.044 \\
    $\pi_{011}$ &       & 0.086 & 0.019 &       & 0.087 & 0.019 &       & 0.087 & 0.019 &       & 0.086 & 0.019 &       & 0.085 & 0.019 &       & 0.085 & 0.019 \\
    $\pi_{111}$ &       & 0.292 & 0.045 &       & 0.294 & 0.044 &       & 0.292 & 0.042 &       & 0.291 & 0.043 &       & 0.304 & 0.040 &       & 0.303 & 0.040 \\
    $\pi_{100}$ &       & 0.024 & 0.006 &       & 0.023 & 0.006 &       & 0.024 & 0.006 &       & 0.024 & 0.006 &       & 0.024 & 0.006 &       & 0.024 & 0.006 \\
    $\pi_{010}$ &       & 0.054 & 0.008 &       & 0.050 & 0.006 &       & 0.052 & 0.007 &       & 0.053 & 0.008 &       & 0.051 & 0.007 &       & 0.051 & 0.007 \\
   $\pi_{110}$ &       & 0.034 & 0.006 &       & 0.033 & 0.005 &       & 0.032 & 0.005 &       & 0.033 & 0.005 &       & 0.034 & 0.006 &       & 0.034 & 0.006 \\
     $\pi_{1\cdot1}$ &       & 0.383 & 0.045 &       & 0.382 & 0.044 &       & 0.378 & 0.041 &       & 0.381 & 0.042 &       & 0.395 & 0.050 &       & 0.396 & 0.050 \\
   $\pi_{\cdot11}$ &       & 0.378 & 0.054 &       & 0.381 & 0.053 &       & 0.379 & 0.053 &       & 0.377 & 0.054 &       & 0.389 & 0.046 &       & 0.388 & 0.046 \\
    $\pi_{1\cdot0}$ &       & 0.058 & 0.008 &       & 0.056 & 0.008 &       & 0.055 & 0.008 &       & 0.056 & 0.008 &       & 0.058 & 0.009 &       & 0.058 & 0.009 \\
   $\pi_{\cdot10}$ &       & 0.088 & 0.012 &       & 0.083 & 0.010 &       & 0.084 & 0.011 &       & 0.086 & 0.011 &       & 0.085 & 0.011 &       & 0.085 & 0.011 \\
    $\g_{101}$ &       & 0.186 & 0.120 &       & 0.180 & 0.115 &       & 0.172 & 0.107 &       & 0.178 & 0.113 &       & 0.175 & 0.111 &       & 0.180 & 0.115 \\
    $\g_{011}$ &       & 0.016 & 0.020 &       & 0.016 & 0.020 &       & 0.016 & 0.019 &       & 0.015 & 0.019 &       & 0.017 & 0.021 &       & 0.017 & 0.020 \\
    $\g_{111}$ &       & 0.066 & 0.039 &       & 0.068 & 0.038 &       & 0.064 & 0.035 &       & 0.067 & 0.036 &       & 0.048 & 0.029 &       & 0.048 & 0.029 \\
    $\g_{100}$ &       & 0.015 & 0.008 &       & 0.015 & 0.008 &       & 0.015 & 0.008 &       & 0.015 & 0.008 &       & 0.014 & 0.008 &       & 0.014 & 0.008 \\
    $\g_{010}$ &       & 0.011 & 0.006 &       & 0.008 & 0.003 &       & 0.011 & 0.005 &       & 0.012 & 0.006 &       & 0.008 & 0.003 &       & 0.008 & 0.003 \\
    $\g_{110}$ &       & 0.010 & 0.006 &       & 0.009 & 0.005 &       & 0.009 & 0.005 &       & 0.009 & 0.005 &       & 0.010 & 0.006 &       & 0.010 & 0.006 \\
    $\tau_{101,011}$ &       & -0.95 & -     &       & -0.95 & -     &       & -0.819 & 0.406 &       & -0.95 & -     &       & -0.95 & -     &       & {-0.95} & {-} \\
    $\tau_{011,111}$ &       & 0.300 & 0.324 &       & 0.310 & 0.307 &       & 0.440 & 0.288 &       & 0.412 & 0.275 &       & 0.000 & 0.193 &       & 0.000 & 0.165 \\
    $\tau_{111,100}$ &       & 0.197 & 0.275 &       & 0.128 & 0.297 &       & -0.015 & 0.277 &       & 0.006 & 0.285 &       & 0.331 & 0.138 &       & 0.331 & 0.136 \\
    $\tau_{100,010}$ &       & -0.029 & 0.219 &       & 0.051 & 0.229 &       & -0.100 & 0.216 &       & -0.053 & 0.196 &       & 0.093 & 0.167 &       & 0.099 & 0.169 \\
    $\tau_{010,110}$ &       & 0.544 & 0.199 &       & 0.607 & 0.196 &       & 0.329 & 0.239 &       & 0.382 & 0.250 &       & 0.618 & 0.258 &       & 0.622 & 0.256 \\

    $-\log L$ &       & \multicolumn{2}{c}{3191.80} &       & \multicolumn{2}{c}{3191.31} &       & \multicolumn{2}{c}{3192.64} &       & \multicolumn{2}{c}{3192.67} &       & \multicolumn{2}{c}{3190.60} &       & \multicolumn{2}{c}{3190.62} \\
    \bottomrule
    \end{tabular}
  \end{small}
\begin{flushleft}
\begin{footnotesize}
$
\mbox{Cln}\{\omega_1^\circ,\omega_2^\circ\}=\left\{
\begin{array}{lcc}
\mbox{Clayton rotated by} \, \omega_1^\circ &\mbox{if} & \tau>0\\
\mbox{Clayton rotated by} \, \omega_2^\circ &\mbox{if} & \tau\leq 0\\
\end{array}\right.
$.
\end{footnotesize}  
\end{flushleft} 
\end{table}

In Table \ref{app-table} we present the results from   the  decomposition of the vine copula density in (\ref{vine-density}), as a different decompositions led to similar results 
due to the small sample size. This is consistent with our previous studies on vine CMMs  \citep{Nikoloulopoulos2015c,Nikoloulopoulos-2018-3dmeta-NE,Nikoloulopoulos-2018-4dmeta-NE}. 
To find the model that provides the best  fit  we don't use goodness-of-fit procedures, but we rather  use the log-likelihood at the maximum likelihood estimate
as a rough diagnostic measure for goodness of fit between the models. 
The goodness-of-fit procedures involve a global distance measure between the model-based and empirical distribution, hence they might not be sensitive to tail behaviours and are not diagnostic in the sense of suggesting improved parametric models in the case of small $p$-values \citep{joe2014}.  For vine copulas, \cite{Dissmann-etal-2013-csda} found that pair-copula selection based on likelihood  seem to be better than using bivariate goodness-of-fit tests.  A larger  likelihood  value indicates a  model that better approximates both  the dependence structure of the data and the strength of dependence in the tails.

 The log-likelihoods showed that a  multinomial truncated D-vine CMM with beta margins and $$
\mbox{Cln}\{0^\circ,90^\circ\}=\left\{
\begin{array}{lcc}
\mbox{Clayton rotated by} \, 0^\circ &\mbox{if} & \tau>0\\
\mbox{Clayton rotated by} \, 90^\circ &\mbox{if} & \tau\leq0\\
\end{array}\right.
$$ bivariate copulas provides the best fit.  Note that as  there exists counter-monotonic  dependence  among   $X_{101}$ and $X_{011}$ ($\tau_{101,011}=-0.95$), this   model
coincides with the model with 
$$
\mbox{Cln}\{0^\circ,270^\circ\}=\left\{
\begin{array}{lcc}
\mbox{Clayton rotated by} \, 0^\circ &\mbox{if} & \tau>0\\
\mbox{Clayton rotated by} \, 270^\circ &\mbox{if} & \tau\leq0\\
\end{array}\right.
$$
bivariate copulas and beta margins as both the Clayton copula rotated by $90^\circ$ and the Clayton copula rotated by $270^\circ$ go to their limiting case the  counter-monotonic copula.

It is revealed that a multinomial truncated D-vine CMM with the vector of probabilities of each combination of tests results in diseased and non-diseased patients  on the original scale provides better  fit than the multinomial GLMM, which models the vector of probabilities of each combination of tests results in diseased and non-diseased patients on a transformed scale. The improvement over the multinomial GLMM is small in terms of the likelihood principle, but for a sample size such as $N=11$, $-3190.60-(-3192.90)=2.3$ units log-likelihood difference is sufficient.

\begin{figure}[!h]
\caption{\label{SROCs} SROC curves  with a predictive  region and summary operating points  (a pair of the model-based joint true positive rate  and joint false positive rate)  from the fitted multinomial truncated  D-vine copula mixed models with beta margins along with the study estimates. } 
\begin{center}
\begin{tabular}{|cc|}
\hline
BVN  &  Frank \\\hline
\includegraphics[width=0.45\textwidth]{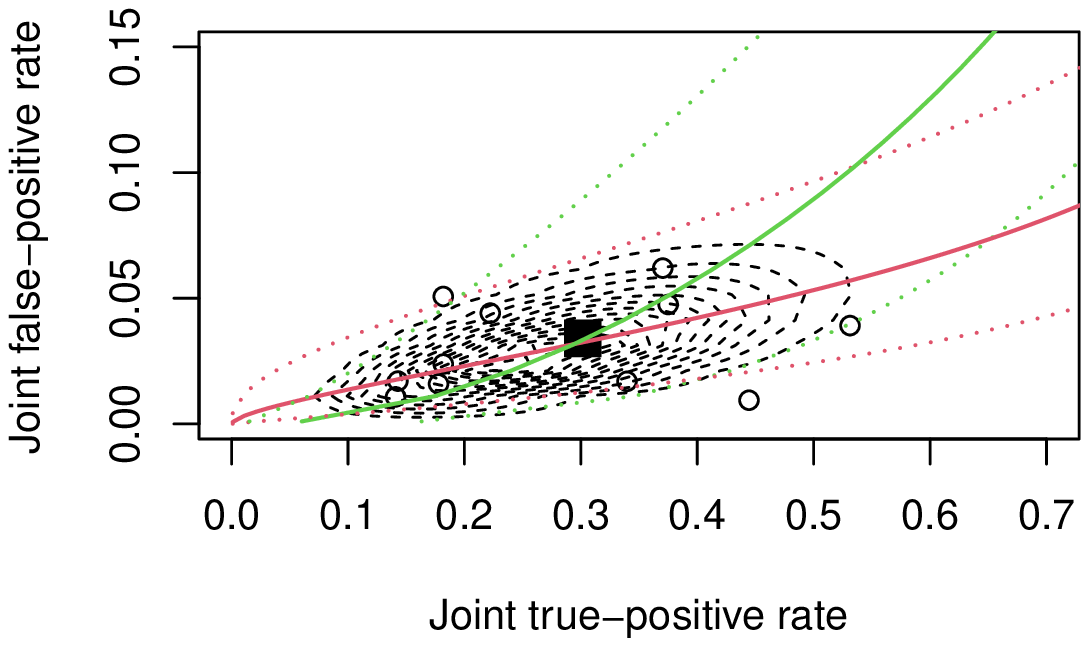}&
\includegraphics[width=0.45\textwidth]{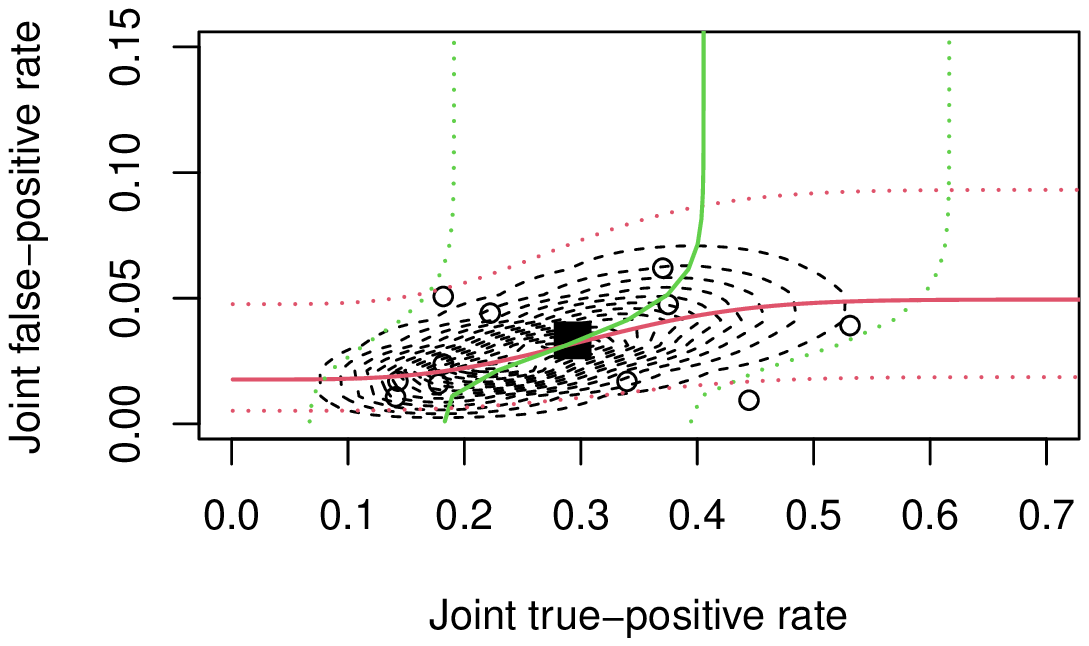} \\\hline
 Clayton &Clayton by $180^\circ$ \\\hline

\includegraphics[width=0.45\textwidth]{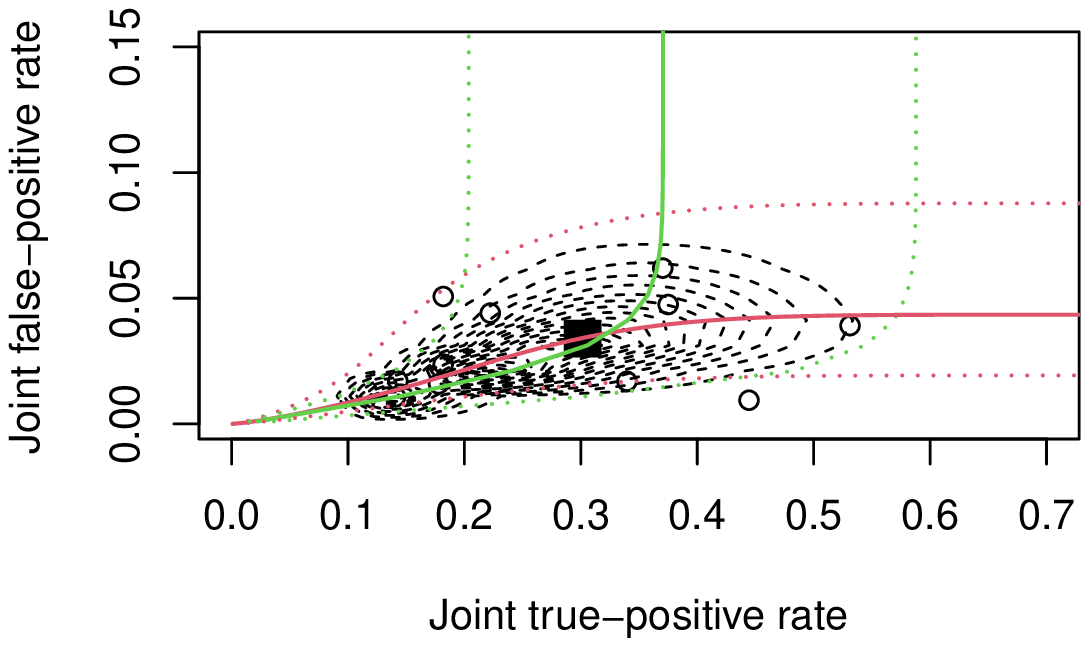} &\includegraphics[width=0.45\textwidth]{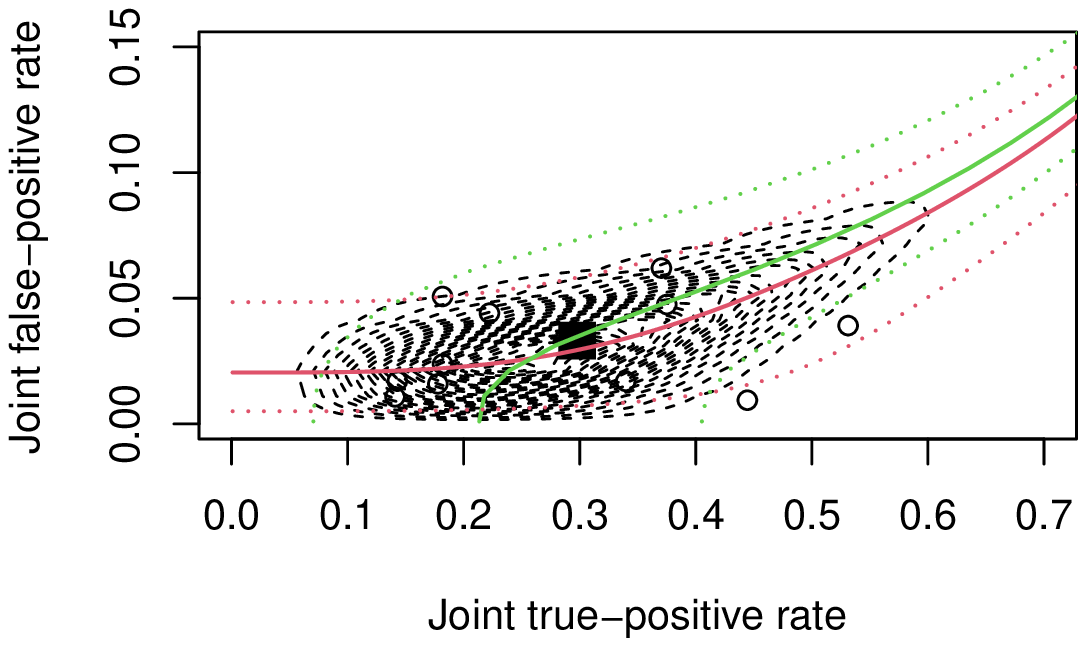}\\
\hline
\end{tabular}

\begin{flushleft}
     \begin{footnotesize}
 $\blacksquare$: summary point; $\circ$: study estimate; 
red and green lines represent the quantile  regression curves $x_{111}:=\widetilde{x}_{111}(x_{110},q)$ and $x_{110}:=\widetilde{x}_{110}(x_{111},q)$, respectively; for $q=0.5$ solid lines and for $q\in\{0.01,0.99\}$ dotted lines (confidence region);    
    \end{footnotesize}
    \end{flushleft}

\begin{footnotesize}

\end{footnotesize}

\end{center}
\end{figure}

 The resultant  meta-analytic estimates of TPR and FPR show that the shortened  humerus marker is better compared with shortened femur. The shortened  humerus marker has both better sensitivity (TPR) and specificity (1-FPR). This is not apparent from the estimated meta-analytic parameters of TPR and FPR for each test under the multinomial GLMM.  
From the Kendall's tau estimates there is strong evidence of dependence between the two diagnostic tests. The fact that the best-fitting bivariate copulas are Clayton reveals that there exists lower tail dependence among the latent vector of probabilities of each combination of tests results in diseased and non-diseased patients. Figure \ref{SROCs} demonstrates the SROC curves  with a confidence  region and summary operating points (a pair of the model-based joint TPR  and joint FPR; shown by the black square) from all the  multinomial truncated  D-vine CMMs with beta margins,  along with the study estimates (shown by the circles in Figure  \ref{SROCs}).  Sharper corners in the  predictive region indicate tail dependence. 
For all the graphs the joint TPR and joint FPR at study $i$  (point estimates) have been calculated as 
$$\frac{y_{i111}}{y_{i++1} }\quad \mbox{and} \quad  \frac{y_{i110}}{y_{i++0}},$$ 
\noindent respectively, and the estimated parameters by refitting the models using the permutation in Section \ref{SROC-section}. The estimated Kendall's $\tau$ association between $X_{111}$ and $X_{110}$ is roughly $\hat\tau_{111,110}=0.45$ from all fitted copulas, but the shapes and regions of the SROCs are distinct as  parametric  bivariate copulas have varying tail behaviour. 
The predictive region from the best fitted  copula (Clayton) has a sharper corner at the lower tail,  
as the  Clayton copula  has lower  tail dependence.

It is  also apparent that the estimates of the meta-analytic parameters  of interest from the multinomial truncated vine CMMs with beta margins  differentiate from the ones  from the multinomial truncated vine CMMs with normal margins.  This is consistent  with the  simulation results and conclusions  in Section \ref{simulations-section}.
The main parameters of interest 
$\pi_{1\cdot t},\pi_{\cdot 1t}$ and $\pi_{11t}$
are biased  when  the univariate random effects are misspecified. 
Our general model can allow both normal and beta margins, i.e., it is not restricted to normal margins as the multinomial GLMM. 

\section{\label{discussion}Discussion}

We have proposed a multinomial truncated D-vine CMM for joint meta-analysis and comparison of  two diagnostic tests.  Our parsimonious model includes the multinomial GLMM \citep{trikalinos-etal-2014-rsm} as a special case, but can also operate on the original scale of the latent probabilities of each combination of test results in diseased and non-diseased patients. It provides an improvement over the multinomial GLMM as  the random effects distribution is expressed via a vine copula that allows for flexible dependence modelling, different from assuming simple linear correlation structures and  normality. Vine copulas, by choosing bivariate  copulas appropriately,  can have a flexible range of lower/upper tail dependence \citep{joeetal10}. The multinomial truncated D-vine CMM allows for selection of parametric bivariate copulas and univariate margins independently among a variety of parametric families. Hence, the latent probabilities of each combination of test results in diseased and non-diseased patients can be modelled on the original proportions scale and can be tail dependent.  

In an era of evidence-based medicine, decision makers need  procedures, such as the SROC curves, to make predictions.  Therefore, for the multinomial  truncated D-vine CMM with beta margins, we  derived  the associated  SROC curves.   The SROC curves essentially show the effect of different model  assumptions such as the choice of parametric bivariate copula and its tail dependence properties, because they are inferences that depend on the joint distribution.
Our proposed model  with normal margins or the multinomial GLMM  \citep{trikalinos-etal-2014-rsm}  cannot be used to produce the SROC curves, since the  latent proportions are modelled on a transformed scale via the multinomial logit link.

We have proposed an efficient  ML estimation technique based on dependent  Gauss-Legendre quadrature points that have a truncated D-vine copula distribution.  
 We use the notion of a truncated at level 1 vine copula that leads to a substantial reduction  of the dependence parameters.  This is extremely useful for estimation purposes  given the typical small sample sizes in meta-analysis of diagnostic test accuracy studies.  
\cite{trikalinos-etal-2014-rsm} estimated the multinomial GLMM using MCMC methods in the Bayesian framework and acknowledged that optimizing the likelihood for joint meta-analysis is non-trivial, because it involves calculating complicated integrals numerically.
Our numerical method  that is based on dependent  Gauss-Legendre quadrature points that have a truncated vine copula distribution successively computes the 6-dimensional integrals in  sextuple sums over the dependent quadrature points and weights.

Authors of primary studies of diagnostic accuracy that assess two tests with paired designs where each test is applied to the same patients  should report the  data as separate $4\times 2$ tables as in Table \ref{4times2}. 
Comparative accuracy studies should rightly be designed so that patients receive each test in order to reduce biases and ensure the clinical relevance of the resulting inferences \citep{trikalinos-etal-2014-rsm}. 
Nevertheless, the proposed model or the multinomial GLMM \citep{trikalinos-etal-2014-rsm} that both consider the case the test results are cross-classified  cannot be extended  to compare  the accuracy of more than two tests as the number of model parameters increase rapidly. For example one needs $2(2^K-1)$ parameters, where $K$ is the number of tests, to only model the  probabilities of each combination of tests results in diseased and non-diseased patients. \cite{Nikoloulopoulos2020-factorREMADA}, without using the information on the agreement between the  tests,  
 proposed an one-factor CMM that can be used for conducting meta-analysis of  comparative accuracy studies with  three or more tests. 

\section*{Software}
{\tt R} functions to implement the multinomial truncated D-vine CMM for meta-analysis of multiple diagnostic tests will be part of the next major release of the  {\tt  R} package {\tt  CopulaREMADA}  \citep{Nikoloulopoulos-2018-CopulaREMADA}.

\section*{Acknowledgements}
 The simulations presented in this paper were carried out on the High Performance Computing Cluster supported by the Research and Specialist Computing Support service at the University of East Anglia.


\end{document}